\documentclass[a4paper, amsfonts, amssymb, amsmath, reprint, showkeys, nofootinbib]{revtex4-1}
\usepackage[english]{babel}
\usepackage[utf8]{inputenc}
\usepackage{xcolor}
\usepackage{upgreek}
\usepackage{amsthm}
\usepackage{mathtools}
\usepackage{physics}
\usepackage{xcolor}
\usepackage{graphicx}
\usepackage[left=20mm,right=20mm,top=30mm,columnsep=15pt]{geometry} 
\usepackage{adjustbox}
\usepackage{placeins}
\usepackage[T1]{fontenc}
\usepackage{lipsum}
\usepackage{csquotes}
\usepackage{float}

\usepackage[pdftex, pdftitle={Article}, pdfauthor={Author}]{hyperref} 
\begin{document}
\title{Cavity-enhanced single artificial atoms in silicon}

\author{Valeria~Saggio$^{1,*}$, Carlos~Errando-Herranz$^{1,2,*}$, Samuel~Gyger$^{1,3}$, Christopher~Panuski$^1$, Mihika~Prabhu$^1$, Lorenzo~De~Santis$^{1,4}$, Ian~Christen$^1$, Dalia~Ornelas-Huerta$^1$, Hamza~Raniwala$^1$, Connor~Gerlach$^1$, Marco~Colangelo$^1$, and~Dirk~Englund$^{1,}$}
\email{vsaggio@mit.edu, carloseh@mit.edu, englund@mit.edu}

\address{\vspace{0.1cm} $^1$Massachusetts Institute of Technology, Cambridge, USA \\$^2$University of M\"unster, M\"unster, Germany \\$^3$KTH Royal Institute of Technology, Stockholm, Sweden \\$^4$QuTech, Delft University of Technology, Delft, Netherlands} 


\begin{abstract}
Artificial atoms in solids are leading candidates for quantum networks~\cite{bhaskar_experimental_2020, pompili_realization_2021}, scalable quantum computing~\cite{childress_diamond_2013, choi_percolation-based_2019-1}, and sensing~\cite{degen_quantum_2017}, as they combine long-lived spins with mobile and robust photonic qubits. 
The central requirements for the spin-photon interface at the heart of these systems are long spin coherence times and efficient spin-photon coupling at telecommunication wavelengths.
Artificial atoms in silicon~\cite{redjem_single_2020, higginbottom_optical_2022} have a unique potential to combine the long coherence times of spins in silicon~\cite{saeedi_room-temperature_2013-1} with telecommunication wavelength photons in the world's most advanced microelectronics and photonics platform.
However, a current bottleneck is the naturally weak emission rate of artificial atoms.
An open challenge is to enhance this interaction via coupling to an optical cavity.
Here, we demonstrate cavity-enhanced single artificial atoms at telecommunication wavelengths in silicon. 
We optimize photonic crystal cavities via inverse design and show controllable cavity-coupling of single G-centers in the telecommunications O-band. Our results illustrate the potential to achieve a deterministic spin-photon interface in silicon at telecommunication wavelengths, paving the way for scalable quantum information processing.
\end{abstract}

\maketitle

\section{Introduction}
Artificial atoms employ controlled coherent spin-photon interfaces to transfer quantum states between long-lived stationary spins and flying photons.
Three central requirements for a scalable spin-photon interface are a long spin coherence time, efficient spin-photon coupling, and operation at telecommunication wavelengths~\cite{ruf_quantum_2021}.
However, current material platforms fail to meet these requirements at once~\cite{zaporski_ideal_2023, pompili_realization_2021, bhaskar_experimental_2020}.

A central challenge is to enhance the naturally weak coherent radiative emission rate while suppressing other excited-state decoherence processes. 
In particular, the modified local density of optical states in a cavity can increase the radiative emission fraction $\beta$ into a desired mode while suppressing emission into other modes:
\begin{equation} \label{eq:beta}
\beta = \frac{(1+F_\text{P}) \gamma_\text{R}}{(1+F_\text{P})\gamma_\text{R} + \gamma_0},
\end{equation}

where $\gamma_R$ is the radiative rate for the transition of interest and $\gamma_0$ encompasses the rates for other radiative and non-radiative transitions.
$F_\text{P}$ is the cavity Purcell factor, which in the case of perfect cavity-atom coupling is defined as~\cite{purcell_spontaneous_1995}
\begin{equation} \label{eq:purcellmain}
F_\text{P}=\frac{3\lambda^3}{4\pi^2}\frac{Q}{V},
\end{equation}
with $\lambda$ the wavelength in the material, $Q$ the quality factor, and $V$ the effective mode volume of the cavity. To enable efficient collection of the light emitted from the cavity, it is 
required that the cavity far-field emission is matched to the optical mode of interest --- such as the mode of an optical fiber. This light collection is quantified using the coupling efficiency $\eta$. 
The net collection efficiency $\beta\eta$ defines the performance of a quantum network built with such devices.
Accommodating both high $Q/V$ and high $\eta$ in a single device is a nontrivial design challenge that depends largely on the materials, the fabrication, and the operation wavelengths of the artificial atom of choice.
These challenges have so far resulted in weak and small-scale spin-photon coupling for current leading artificial atom platforms~\cite{ruf_quantum_2021}. 

\begin{figure*}[htbp]
  \centering
  \includegraphics[width=\linewidth]{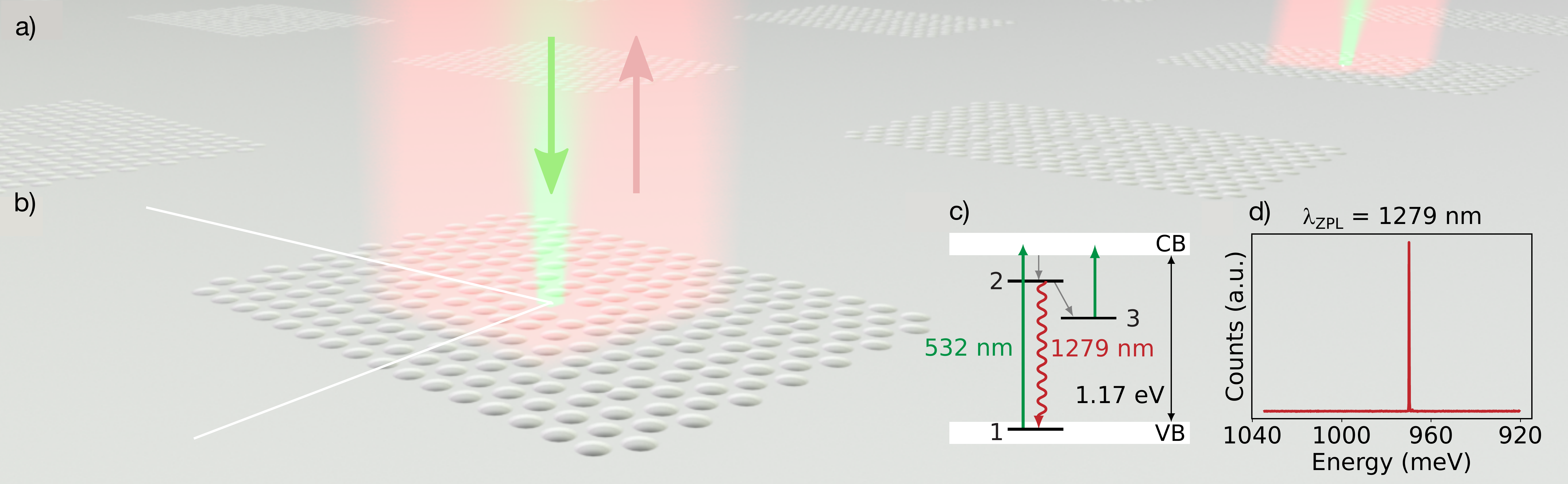}
\caption{\textbf{Description of the system.} a) Illustration of the system under study, consisting of optimized 2D photonic crystal cavities coupled to single G-centers in silicon. b) The G-center consists of two substitutional carbon atoms (blue spheres) and a silicon interstitial atom (gray spheres). c) The O-band radiative transition occurs between singlet states 1 and 2 and can be excited via above-band excitation. The system features an additional triplet state (3). d) Spectrum showing the photoluminescence (PL) of a single cavity-coupled G-center with a zero phonon line (ZPL) around 1279~nm.}
\label{fig:motivation}
\end{figure*}

Recently, there has been a resurgence of interest in silicon as a host material for single artificial atoms operating in the telecommunication bands~\cite{bergeron_silicon-integrated_2020, baron_detection_2021, durand_broad_2021, higginbottom_optical_2022}.
Initial reports have focused on the G-center~\cite{hollenbach_engineering_2020, redjem_single_2020}, the T-center~\cite{higginbottom_optical_2022, deabreu2022waveguide}, and the W-center~\cite{baron_detection_2021}, and include the first optical observation of an isolated spin in silicon~\cite{higginbottom_optical_2022}, and the first isolation of single artificial atoms in silicon waveguides and their spectral programming~\cite{prabhu_individually_2022}.
These demonstrations, combined with the experimentally reported 30-min-long spin coherence times in ionized donors in $^{28}$Si~\cite{saeedi_room-temperature_2013-1} and the success of silicon microelectronics and photonics~\cite{vivien_handbook_2016}, make this technology compelling for large-scale quantum information processing.
However, although high $Q/V$ and $\eta$ optical cavities have recently been shown in silicon at a large scale~\cite{panuski2022full}, their coupling to single artificial atoms remains a challenge.

Here, we report on the inverse design of high $Q/V$, $\eta$-optimized photonic crystal cavities, and demonstrate cavity-enhanced interaction of light with single artificial atoms at telecommunication wavelengths in silicon.

\section{Results}
Our device, illustrated in Fig.~\ref{fig:motivation}a, consists of single G-centers coupled to inverse-designed 2D photonic crystal cavities. 
The G-center is a quantum emitter formed by two substitutional carbon atoms and a silicon interstitial (Fig.~\ref{fig:motivation}b), and features a zero phonon line (ZPL) transition at 970~meV (1279~nm) in the telecommunications O-band along with a spin triplet metastable state~\cite{udvarhelyi_identification_2021} (Figs.~\ref{fig:motivation}c, d).

\begin{figure*}[htbp]
  {\centering
  \includegraphics[width=\linewidth]{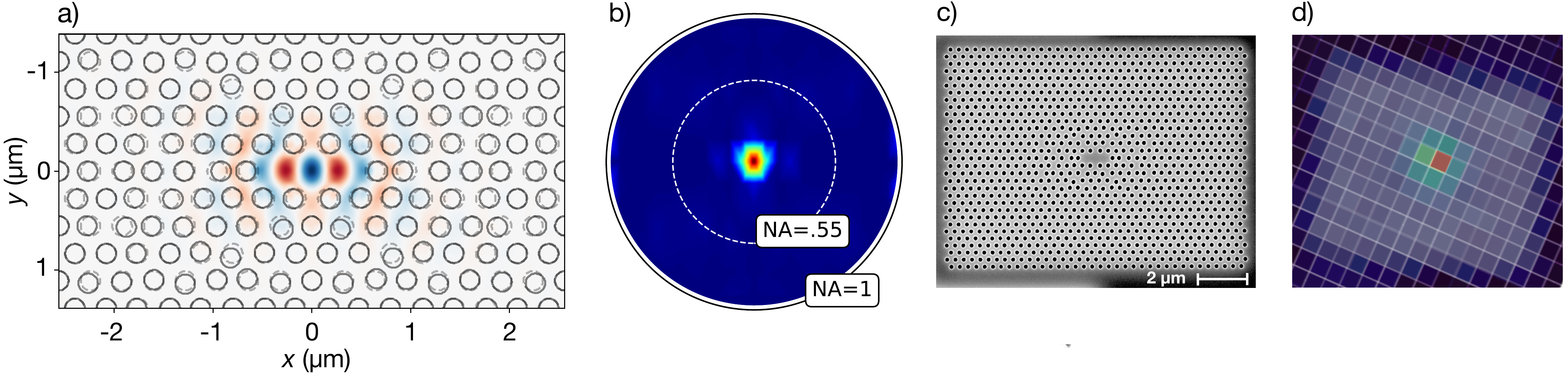}}
\caption{\textbf{Optimized cavities.} Simulated a) near-field electric field amplitude and b) scattered far-field power for the 2D photonic crystal cavity design used in this work. c) Scanning electron micrograph showing one of our optimized photonic crystal cavities. d) Confocal PL 2D scan showing the reflectivity map (blue scale) overlaid with a PL map (red and green) of the same area, highlighting the PL emission from the cavity center.}
\label{fig:cav_ff}
\end{figure*}

Our cavities were designed following our previous work~\cite{panuski2022full} to simultaneously achieve a target $Q/V$ while optimizing for vertical coupling $\eta$ by matching the emission to a narrow numerical aperture in the O-band.
Fig.~\ref{fig:cav_ff} shows one of our cavity designs, including its near-field cavity mode (Fig.~\ref{fig:cav_ff}a) and its far-field scattering profile (Fig.~\ref{fig:cav_ff}b), with more than 70\% of the emitted power simulated to radiate into an objective NA of 0.55.
More information on the optimization can be found in Methods, and the results of the cavity optimization in SI Section~\ref{sec:optimization} and Fig.~\ref{sfig:optimization}.
The fabrication of our device follows our previous work~\cite{prabhu_individually_2022} with the addition of an underetch step, and is described in Methods.

The measurements on our device were performed with a setup consisting of a home-built cryogenic confocal microscope featuring temperature and CO$_2$ gas control, and optimized for visible light excitation and infrared collection into a single-mode fiber (details in SI Section~\ref{sec:setup}).
A scanning electron micrograph of a representative cavity in our chip is shown in Fig.~\ref{fig:cav_ff}c. Its reflectivity was characterized in cross-polarization~\cite{altug_polarization_2005, panuski2022full,deabreu2022waveguide} (details in SI Section~\ref{sec:cross}). Fig.~\ref{fig:cav_ff}d shows a photoluminescence (PL) 2D scan of one of our systems, where the cavity-coupled artificial atom is evidenced by the color-labeled IR emission in the cavity center upon excitation with green light. The spectral signature of the PL, shown in Fig.~\ref{fig:motivation}d, features a ZPL centered at around 1279~nm and thus aligns
with the previously reported G-center ZPLs~\cite{hollenbach_engineering_2020, baron_detection_2021, prabhu_individually_2022}.

For the cavities of interest, we measure quality factors of $\sim3700$ and 2100, and center wavelengths around 1279~nm, near the G-center ZPL.

\begin{figure*}
  \centering
  \includegraphics[width=\linewidth]{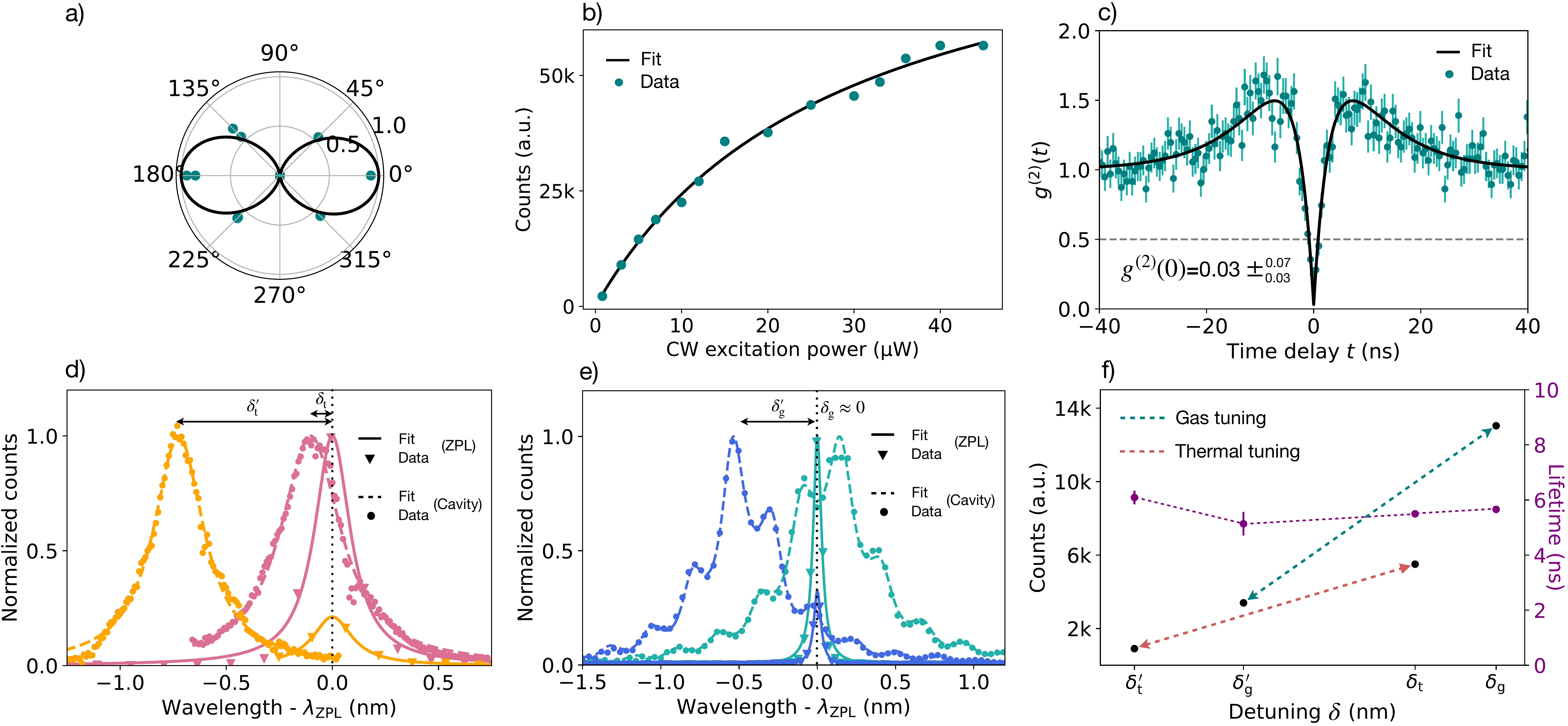}
\caption{\textbf{Cavity-enhanced single-photon emission.} a) Polarization plot of the PL emission from our system, which matches that of an electric dipole.  
b) PL counts at increasing CW excitation powers. Saturation matching that of a two-level system is observed. 
c) The second-order autocorrelation function, which yields $g^{(2)}(0) \approx 0$, demonstrating high-purity single-photon emission. d) Cavity-atom coupling is demonstrated by changing the sample temperature from 4~K (pink) to 24~K (orange) to spectrally tune a cavity with respect to the G-center ZPL, which results in a reduction of the PL magnitude. e) This effect is confirmed with a second cavity-atom system on the same chip before (green) and after (blue) gas detuning. f) PL counts and lifetimes for both systems under the measured cavity-atom detuning magnitudes.}
\label{fig:coupling}
\end{figure*}

Fig.~\ref{fig:coupling} shows the experimental results confirming the presence of single artificial atoms in our photonic crystal cavities. The coupling between the atom and the cavity results in an enhancement of the atom single-photon emission. We observe linearly polarized PL emission (Fig.~\ref{fig:coupling}a) from our system, indicative of coupling through the expected transverse electric cavity mode. 
Measuring the PL saturation under increasing excitation power yields that of a two-level emitter model (see Fig.~\ref{fig:coupling}b and SI Section~\ref{sec:pulsed}).
We further confirmed the addressing of a single artificial atom by demonstrating single-photon emission via a Hanbury-Brown-Twiss (HBT) experiment.
Our second-order autocorrelation results (Fig.~\ref{fig:coupling}c) show excellent antibunching with a fitted $g^{(2)}(0)$ value of $0.03 \substack{+0.07 \\ -0.03}$ without background correction (details in SI Section~\ref{sec:g2}). 
This value is nearly an order of magnitude lower than the rest of the literature (see SI Table~\ref{tab:comparison}), and indicates high-purity single-photon emission.
The bunching near $\pm 10$~ns delay conforms with the presence of a third dark state, which has been attributed to a metastable triplet state~\cite{udvarhelyi_identification_2021}.
These measurements demonstrate the presence of a single G-center in our cavity.

To confirm the cavity enhancement of our single artificial atom, we detuned the cavity resonance wavelength from the G-center ZPL using two different methods, i.e. thermal and gas tuning.
In the thermal tuning experiments, starting with a cryostat temperature of 4~K, we brought the temperature up to 24~K and consequently shifted the cavity away from our G-center ZPL. This effect is visible in Fig.~\ref{fig:coupling}d, where the pink (orange) curves show the cavity and ZPL profiles before (after) the temperature increase. Different detunings $\delta_{\mathrm{t}}$ and $\delta_{\mathrm{t}}^{\prime}$, defined as the difference between the cavity and ZPL wavelengths, are therefore achieved at 4~K and 24~K, respectively.  
While a significant cavity wavelength shift occurs, we also observe a much smaller ZPL shift, not shown in the figure. Therefore, this plot shall be indicative only of the relative shift between the cavity and ZPL wavelength. A more detailed discussion about the figure can be found in SI Section~\ref{sec:Tuningofcavity}a. 
We note that temperatures below 30~K have been reported to not affect G-center ensembles~\cite{beaufils_optical_2018, chartrand_highly_2018}.
We observe an intensity enhancement of $6.078\pm0.218$ with a cavity $Q$ of $\sim 3700$ and a mode volume of $V<1(\lambda/n)^3$.

To validate that the intensity enhancement does not originate from the temperature change induced by thermal cavity tuning, we performed additional experiments using gas tuning. 
We injected CO$_2$ gas into the cryogenic sample chamber to coat the cavity with solid CO$_2$, followed by selective gas sublimation using a $532$~nm continuos wave (CW) laser.
A further description of the process can be found in SI Section~\ref{sec:Tuningofcavity}b. 
Analogously to the thermal tuning case, Fig.~\ref{fig:coupling}e shows the cavity reflectivity and G-center ZPL now under gas tuning for two different detunings $\delta_{\mathrm{g}}$ and $\delta_{\mathrm{g}}^{\prime}$. We calculate an enhancement in the PL emission of $3.837\pm0.074$ with a cavity $Q$ of $\sim 2100$ and a mode volume of $V<1(\lambda/n)^3$. Also in this case, we observe a ZPL shift (not shown in the figure). More details are to be found in SI Section~\ref{sec:Tuningofcavity}b.

Ultimately, we measured the excited state lifetime of our artificial atoms under both gas and thermal tuning using a 0.5~ns pulsed laser at 532~nm for all cavity detunings (see SI Section~\ref{sec:pulsed} for details about these measurements). 
Fig.~\ref{fig:coupling}f shows our measured lifetimes and the emission rates for all of our experiments.
We do not observe a statistically significant lifetime modification even under a clear cavity enhancement of the emission rates above 6x.

The quantum efficiency (QE) of any silicon color center is one of the central unanswered questions in the field. 
Recent reports have estimated the QE of a single G-center to be above 1\% from waveguide-coupled counts~\cite{prabhu_individually_2022, komza_indistinguishable_2022}, and $<$~10\% for ensembles coupled to separate cavities~\cite{lefaucher2023cavity}.
To gain a valid estimate of the QE of the G-center, an experiment varying the coupling rate between the same single artificial atom and a cavity was required.
Our measurements allow us to extract such a value.
Using the derivation described in SI Section~\ref{sec:qe}~\cite{lefaucher2023cavity}, the literature value of the Debye-Waller factor $F_\text{DW}=0.15$~\cite{beaufils_optical_2018}, and our measured values of off-resonance lifetime $\tau_\text{off}=6.09\pm0.25$~ns and count rate enhancement for thermal tuning, 
we obtain a quantum efficiency bounded as $\text{QE}~<~18\%$.

\section{Discussion}
We show, for the first time, strong enhancement of the quantum emission of individual artificial atoms coupled to silicon nanocavities.
A central requirement for the scalability of our system is localized spatial and spectral alignment of both many cavities and many atoms to a common global frequency.
The spatial alignment of the cavity and atom can be achieved by making use of the recently reported localized implantation of single G- and W-centers~\cite{hollenbach_wafer-scale_2022}.
Silicon artificial atoms can be spectrally aligned using the recently reported non-volatile optical tuning for G-centers~\cite{prabhu_individually_2022} or methods used in other artificial atom systems such as tuning via electric fields~\cite{anderson_electrical_2019-2}, or mechanical strain~\cite{wan_large-scale_2020-2}.
Cavity tuning via local thermal oxidation of silicon has been achieved on a large scale~\cite{panuski2022full}, and a similar method could be used to align large arrays of cavity-atom systems at room or cryogenic temperatures.
Our approach directly applies to other silicon artificial atoms, such as the T-center, which would enable direct access to a spin outside of a metastable state~\cite{higginbottom_optical_2022}.

A hypothesis was recently raised regarding the possibility of two different physical systems being reported as G-centers~\cite{baron_single_2022}.
We believe that our work provides conclusive evidence for this claim.
Table~\ref{tab:comparison} in SI Section~\ref{sec:comparison} compares the reported experimental results for single G-center labeled artificial atoms in silicon, and shows two clear clusters.
The first group comprises the single emitter reports in Refs.~\cite{hollenbach_engineering_2020, baron_single_2022, prabhu_individually_2022, komza_indistinguishable_2022}, and aligns with G-center ensemble work~\cite{lefaucher2023cavity}. These studies show a ZPL centered around 1279~nm and a narrow inhomogeneous linewidth <~1.1~nm, a QE between 1 and 18\%, and a short excited state lifetime <~10~ns that does not change significantly under Purcell enhancement, confirming the QE magnitude.
The second group comprises Refs.~\cite{redjem_single_2020, redjem_all-silicon_2023}, and features a shorter ZPL centered around 1270~nm and a larger inhomogeneous linewidth of 9.1~nm, a QE$\sim$50\%, and a longer excited state lifetime >~30~ns, which changes significantly under Purcell enhancement and thus qualitatively aligns with the estimated QE.
Our measurements align with the first system, i.e. the originally reported G-centers, and provide the first upper bound for the QE of single G-centers, previously estimated to be between 1\%~\cite{prabhu_individually_2022, komza_indistinguishable_2022} and 10\% for ensembles~\cite{lefaucher2023cavity}.
More information on this comparison can be found in SI Section~\ref{sec:comparison}.
We conclude that our work highlights the need for further theoretical and experimental investigation regarding the creation process and the photophysics of G-center-like artificial atoms in silicon platforms.

\section{Conclusion}
We showed cavity-enhanced single artificial atoms in silicon by integrating single G-centers into inverse-designed photonic crystal nanocavities.
We demonstrated an intensity enhancement of $\sim$6x with a cavity featuring a $Q/V>3700$, which yielded the highest purity single-photon emission for silicon color centers in the literature, and the first bound to the QE of single G-centers of $<18$\%.
Our demonstration lays the groundwork for efficient spin-photon interfaces at telecommunication wavelengths for large-scale quantum information processing in silicon.

\section*{Methods}
\subsection{Sample fabrication}
The fabrication process follows~\cite{redjem_single_2020}, starting from a commercial SOI wafer with $220$~nm silicon on $2$~\textmu m silicon dioxide. 
Cleaved chips from this wafer were implanted with $^{12}$C with a dose of $5\times10^{13}$~ions/cm$^{2}$ and $36$~keV energy, and subsequently annealed at $1000~^{\circ}$C for $20$~s to form G-centers in the silicon layer.
The samples were then processed by a foundry (Applied Nanotools) for electron beam patterning and etching, resulting in through-etched silicon cavities with SiO$_2$ bottom cladding and air as top cladding.
The silicon etching was performed using inductively coupled plasma reactive ion etching with SF$_6$-C$_4$F$_8$ mixed-gas.
As a final step, the samples were under-etched in a $49\%$ solution of hydrofluoric acid for $2$~min and dried using a critical point dryer.

\subsection{Cavity far-field optimization}
\label{methods:cav_ff}

Traditional photonic crystal cavity optimization aims to cancel radiative loss to enhance quality factor $Q$, which also reduces collection efficiency.
To avoid this, we incorporate the far-field collection efficiency $\eta$ to the optimization objective function alongside maximizing $Q$ and minimizing mode volume $V$~\cite{panuski2022full} (Fig.~\ref{sfig:optimization}).
This process is implemented using the open-source package \texttt{Legume}~\cite{Minkov_inverse_2020} which maps the problem of cavity design onto efficient and auto-differentiable guided mode expansion (GME) for gradient-based optimization.

In practice, we observe quality factors much lower than the $Q \sim \mathcal{O}(10^6)$ result expected from both simulation (Fig.~\ref{sfig:optimization}) and previous statistical studies on thousands of photonic crystals designed for $\sim$1550~nm operation under the same optimization method~\cite{panuski2022full}. 
We attribute this disparity to the high carbon doping density used to produce cavity-coupled G-centers with sufficient probability. Reducing the doping density or applying localized doping~\cite{hollenbach_engineering_2020} could play a role in recovering performance closer to intrinsic silicon.
Applying large-scale characterization techniques~\cite{sutula_2022_largescale} to locate ideal emitters and fabricate cavities around these positions could enhance the yield of coupled emitters in the case of reduced doping density.

\section*{Notes}

During the preparation of this manuscript, we became aware of a manuscript reporting on a cavity-coupled single silicon color center~\cite{redjem_all-silicon_2023} and a second manuscript reporting on single rare-earth ions in silicon cavities~\cite{gritsch_purcell_2023}.

\section*{Acknowledgements}
The authors acknowledge Kevin C. Chen, Chao Li, Hugo Larocque and Mohamed ElKabbash for helpful discussions. 
C.E-H. and L.D. acknowledge funding from the European Union’s Horizon 2020 research and innovation program under the Marie Sklodowska-Curie grant agreements No.896401 and 840393.
M.P. acknowledges funding from the National Science Foundation (NSF) Convergence Accelerator Program under grant No.OIA-2040695 and Harvard MURI under grant No.W911NF-15-1-0548.
I.C. acknowledges funding from the National Defense Science and Engineering Graduate (NDSEG) Fellowship Program and NSF award DMR-1747426.
M.C. acknowledges support from MIT Claude E. Shannon award.
D.E. acknowledges support from the NSF RAISE TAQS program.
This material is based on research sponsored by the Air Force Research Laboratory (AFRL), under agreement number FA8750-20-2-1007. 
The U.S. Government is authorized to reproduce and distribute reprints for Governmental purposes notwithstanding any copyright notation thereon.
The views and conclusions contained herein are those of the authors and should not be interpreted as necessarily representing the official policies or endorsements, either expressed or implied, of the Air Force Research Laboratory (AFRL), or the U.S. Government.

\bibliography{Si_artificial_atoms.bib}

\newpage
\onecolumngrid
\appendix
\renewcommand{\thefigure}{S\arabic{figure}}
\setcounter{figure}{0}
\newpage 

\section*{Supplementary Information}

\subsection{Cavity optimization}
\label{sec:optimization}
\begin{figure}[htbp]
  {\centering
  \includegraphics[width=0.4\textwidth]{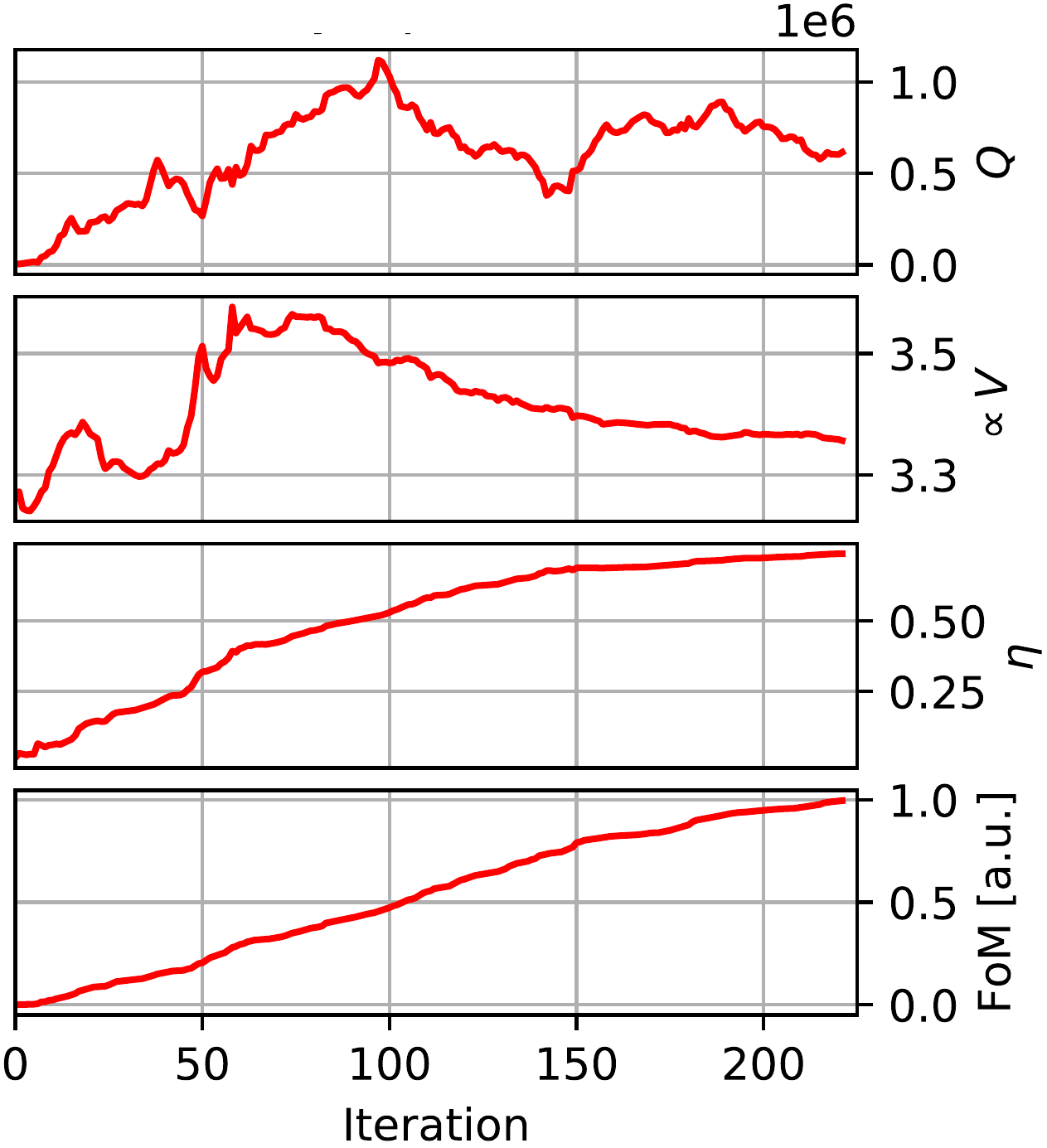}}
\caption{\textbf{2D photonic crystal cavity optimization path.} The course of optimization for one of our designs over 222 iterations, plotting the simulated quality factor $Q$, a result proportional to the mode volume $V$, the scattering efficiency $\eta$ into the 0th Brillouin zone, and the composite figure of merit FoM targeted by the optimization.}
\label{sfig:optimization}
\end{figure}

Figure~\ref{sfig:optimization} shows the course of optimization for the two cavities we employed in our experiments, with the only difference being the hole sizes. Sweeps with several varying parameters (e.g. lattice constant, cavity and hole size) were fabricated in order to target systems with the desired performance.   

\subsection{Measurement setup}
\label{sec:setup}
\begin{figure}
  {\centering
  \includegraphics[width=0.64\textwidth]{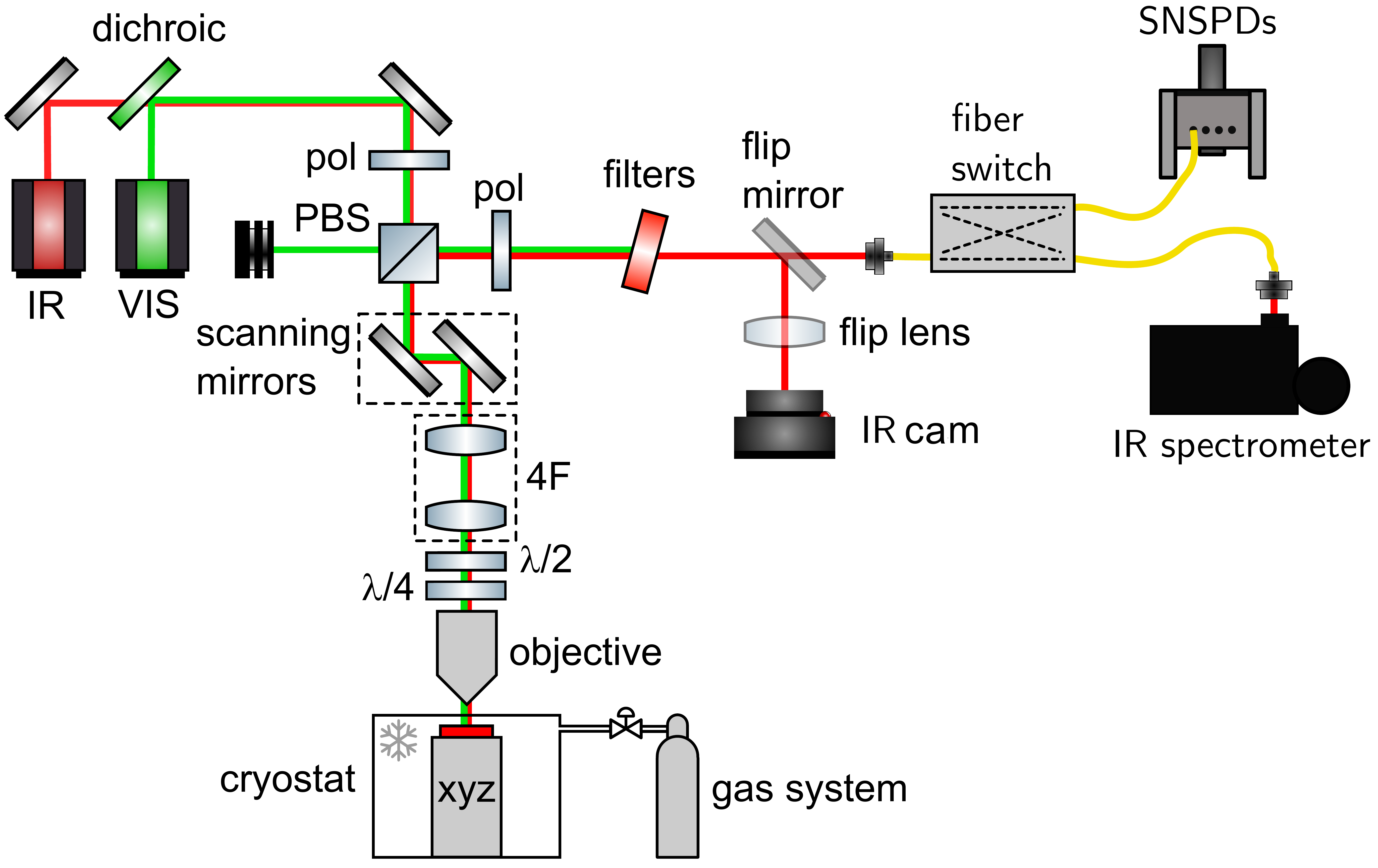}}
\caption{\textbf{Measurement setup schematic.} Our setup consists of a cryogenic confocal microscope. Laser light at visible (532~nm) and infrared (1280~nm) shines through a polarizer and a polarizing beam splitter (PBS), a set of galvanometers (scanning mirrors), a 4F system, and polarization rotation components, into an objective and into the cryostat. The IR PL or cavity reflectivity from the sample reflects off the PBS through a polarizer and a filtering stage into the fiber switch, which leads to SNSPDs or an IR spectrometer. A flip mirror and a lens between the filters and the fiber switch into an IR camera enable the visualization of the image plane, and flipping away the lens provides us with a back focal plane image.}
\label{sfig:m3setup}
\end{figure}

Fig.~\ref{sfig:m3setup} shows a schematic of our measurement setup, which consists of 3 main optical paths: 1) excitation, 2) cryogenic 4F, and 3) collection.

The excitation path combines IR and visible collimated lasers with a dichroic, and routes them via a linear polarizer into a polarizing beam splitter (PBS).
For our characterization, we use two visible lasers and two infrared lasers. 
Our visible lasers consist of a continuous wave Coherent Verdi G5 at 532~nm, and a pulsed laser from NKT Photonics SuperK laser with a maximum repetition rate of 78~MHz and filtered by a bandpass filter centered at 532~nm with a bandwidth of 0.2~nm. 
Our infrared lasers are a tunable CW O-band TSL570C from Santec, and a superluminescent diode S5FC1018S from Thorlabs with a broadband emission centered at 1310~nm. A set of scanning mirrors is placed in the excitation path for precise beam positioning and PL mapping of our cavities. 

The transmitted polarization component is imaged into our cryostat by using a 4F lens system consisting of scanning piezoelectric mirrors, two lenses, and an objective, with preceding IR quarter and half wave plates for polarization rotation.
Our microscope objective is a collar corrected objective LCPLN50XIR from Olympus with a NA of 0.65, and is external to the cryostat. 
Our cryostat is a Montana Instruments system. 
The sample is mounted on a XYZ cryogenic piezoelectric stage (Attocube).

The PL and reflection from our sample are collected via the PBS reflection through an IR polarizer and a filtering station into a fiber switch, which routes the collected light into superconducting nanowire single-photon detectors (SNSPDs) or an IR spectrometer.
Our filtering setup consists of a longpass filter (cutoff wavelength at 1250~nm) and a shortpass filter (cutoff wavelength at 1300~nm). 
Additionally, in several experiments we used a tunable fiber filter from WLPhotonics with FWHM transmission bandwidth of $0.10$~nm.
Our two SNSPDs (NIST) feature detection efficiencies of $21\%$ and $24\%$, and are readout with a Swabian Instruments Timetagger~20. 
Our IR spectrometer consists of a PyLon IR CCD from Princeton Instruments and switchable gratings, one with a density of 300~gr/mm and a 1.2~\textmu m blaze and another with a density of 900~gr/mm and a 1.3~\textmu m blaze, leading to pixel-defined resolutions of 155~pm and 40~pm respectively.
For the second-order autocorrelation measurements we used a fiber beam splitter (Thorlabs TW1300R5F1) after the filtering station.
In addition, we image our sample using a flip mirror before the fiber switch and an InGaAs cooled CCD camera (Allied Vision Goldeye), preceded by a flip lens that allows us to switch between near-field and far-field imaging.

The setup comprises also a gas line equipped with a needle valve and several ball valves to enable controllable gas injection into our cryostat. 

\subsection{Cross-polarization cavity characterization}
\label{sec:cross}
We characterized our cavities via a cross-polarization measurement, as previously reported in Ref.~\cite{altug_polarization_2005}.
The measurement protocol consists of preparing the polarization of the input and output IR beams to be orthogonal and $45^\circ$ rotated with respect to the cavity axis. 
This was achieved by setting the input and output polarizers perpendicular to each other, and by rotating the common half-wave plate to align the fields to the cavity (see schematic in Fig.~\ref{sfig:m3setup}). 
In our setup, we used a PBS for increased polarization extinction.

\subsection{Measurements under pulsed excitation}
\label{sec:pulsed}
All measurements under pulsed excitation were performed with our SuperK laser using a repetition rate of 39~MHz, and varying the power from a minimum of 0.2~$\upmu$W to a maximum of 2~$\upmu$W. 

\begin{figure}
    \centering
    \includegraphics[width=0.83\textwidth]{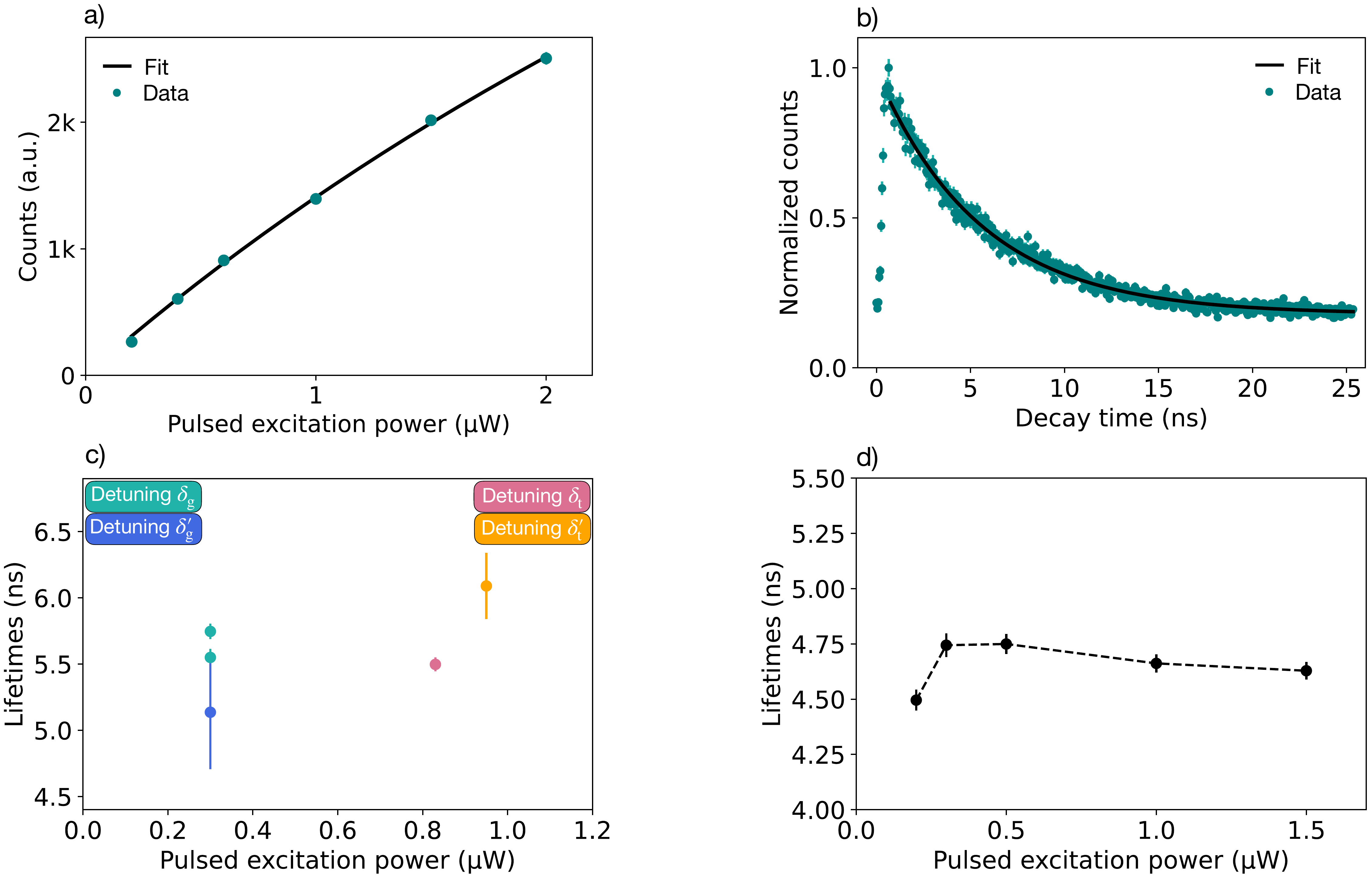}
    \caption{\textbf{Measurements under pulsed excitation}. a) Saturation curve obtained acquiring PL counts while varying the pulsed excitation power. b) Example of a lifetime fit, where a mono-exponential is used to obtain the lifetime from the experimental data. c) Lifetimes displayed both for the gas and thermal tuning cases. In each case, we measured the lifetimes at similar pulsed excitation powers. d) Set of lifetime measurements collected at detuning $\delta_\mathrm{g}$ under possibly different laser conditions.
    }
    \label{pulsed_meas}
\end{figure}

As already discussed in the main text, our artificial atom emission fits well to the characteristic two-level emitter saturation model, which is given by 
\begin{eqnarray}
    I(P)=I_{\infty}\frac{P}{P+P_\text{sat}}.
\end{eqnarray}
Here, $I$ and $P$ are intensity and power, respectively, and $I_{\infty}$ and $P_{\text{sat}}$ are the corresponding saturation values. Fitting our experimental data to this theoretical model, we find $I_{\infty}=(12\pm 2$)~kcounts/min and $P_{\text{sat}}= (7.4 \pm 1.2)$~$\upmu$W. The data and corresponding fit are shown in Fig.~\ref{pulsed_meas}a. The same theoretical model is used to fit the data taken under CW excitation power reported in the main text. In that case (Fig.~\ref{fig:coupling}b in the main text), we obtain $I_{\infty}^{\mathrm{CW}}=(93\pm 5$)~kcounts/min and $P_{\text{sat}}^{\mathrm{CW}}= (28 \pm 3)$~$\upmu$W. 

Fig.~\ref{pulsed_meas}b shows an example of our lifetime data and fits. The plot displays one data set acquired at detuning $\delta_{\mathrm{g}}$ (refer to Fig.~\ref{fig:coupling}e in the main text) with a pulsed excitation power of 0.3~$\upmu$W fitted to a mono-exponential model

\begin{eqnarray}
    f(t) = x\ e^{-\frac{t-y}{\tau}}+z,
    \label{mono}
\end{eqnarray}
with $x$ being a fitting constant for the amplitude, and $y$ and $z$ offsets for the decay time $t$ and $f(t)$, respectively. 
The fit returns a lifetime of $\tau=5.55 \pm 0.06$~ns. All lifetime data sets are fitted using the same model, which returns the lifetime values plotted in Fig.~\ref{pulsed_meas}c for all the detuning values. In both thermal and gas tuning cases, the lifetime does not differ significantly for different detunings. This figure displays the values that are used to estimate the bound on the QE. The single value shown in Fig.~\ref{fig:coupling}f in the main text for detuning $\delta_{\mathrm{g}}$ is the weighted average between the two values reported in Fig.~\ref{pulsed_meas}c. However, for the sake of completeness, we also report additional lifetimes acquired at detuning $\delta_{\mathrm{g}}$ under likely different laser conditions, originating from an error related to our laser control electronics board. The values are shown in Fig.~\ref{pulsed_meas}d. In this case, we measure lifetimes all below $4.75$~ns. As these measurements were taken under possibly different experimental conditions, we decided not to include them in our theoretical analysis. However, they confirm that the lifetime of our emitter remains essentially unchanged with increasing excitation power.

\subsection{Second-order autocorrelation measurements}
\label{sec:g2}
\begin{figure}
  {\centering
  \includegraphics[width=0.95\textwidth]{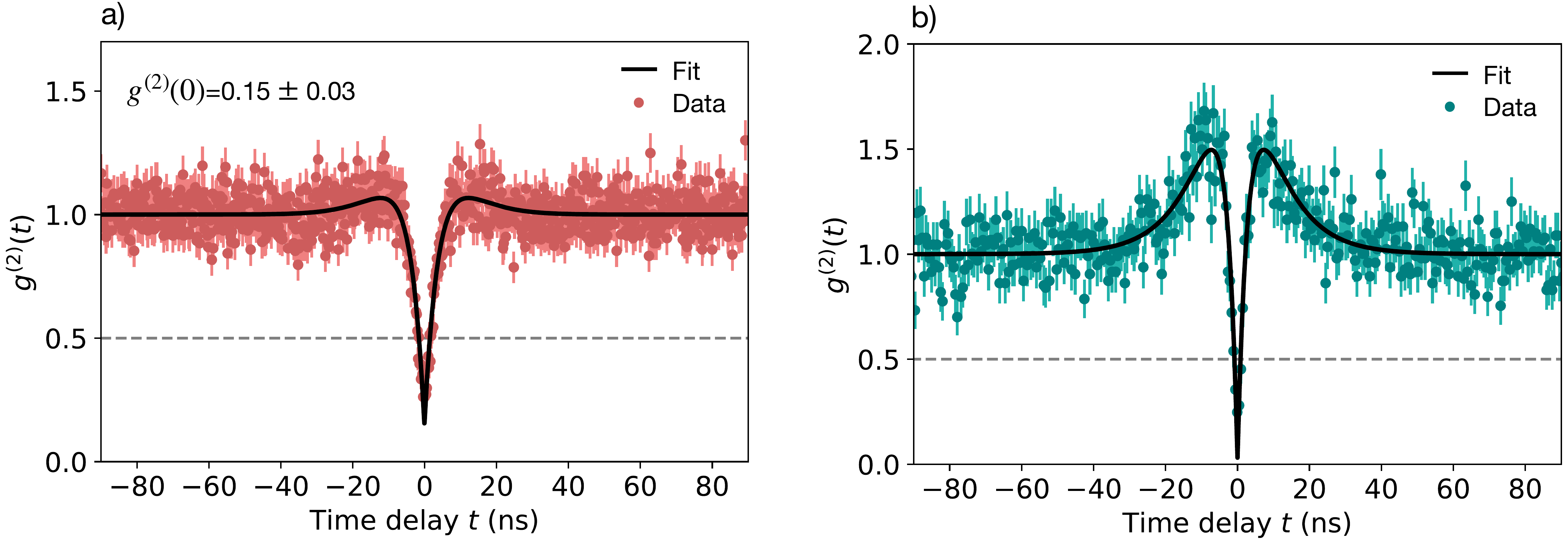}}
\caption{\textbf{Second-order autocorrelation measurements.} a) HBT measurement result from the thermally tuned cavity-atom system, and b) zoomed out version of the gas tuning cavity-atom system shown in Fig.~\ref{fig:coupling}d. Both measurements show clear $g^{(2)}(0)\approx0$, demonstrating single-photon emission.}
\label{sfig:g2s}
\end{figure}

The second-order autocorrelation measurements were performed using a HBT interferometer. 
We excited each of our emitters with a $532$~nm CW pump and sent the generated photons to a fiber beam splitter whose outputs were connected to two SNSPDs, and we then analyzed the coincidence counts at different time delays between the two outputs. 
Fitting our second-order autocorrelation data with a three-level system equation
\begin{equation}
    g^{(2)}(t) = a\left[ 1-(1-b) \left( (1+c) e^{-\frac{\abs{t-t_{\mathrm{shift}}}}{\tau_1}} -c\  e^{-\frac{\abs{t-t_{\mathrm{shift}}}}{\tau_2}}\right) \right],
\end{equation}
with $a$, $b$, and $c$ fitting parameters, $t_\text{shift}$ the offset for the time delay $t$, and $\tau_1$ and $\tau_2$ the lifetimes, we obtain the $g^{(2)}(0)$ value after data normalization as $g^{(2)}(0)=b$ at $t=t_{\mathrm{shift}}=0$. HBT measurement data and fits are shown in Figs.~\ref{sfig:g2s}a and b for the thermal tuning and gas tuning cases, respectively. 
In both cases we obtain a $g^{(2)}(0)$ close to 0, thus confirming genuine single-photon emission. 
In the gas tuning case, we find $g^{(2)}(0)^{\mathrm{g}}= 0.03 \substack{+0.07 \\ -0.03} $, $\tau_1^{\mathrm{g}} = (3.16
\pm 0.5)$~ns and $\tau_2^{\mathrm{g}} = (9.05 \pm 1.3)$~ns. The data was collected filtering the region around the ZPL, and thus reducing the noise contribution coming from elsewhere. 
In the thermal tuning case, we used the longpass and shortpass filters to isolate a $\sim 50$~nm-wide region comprising the ZPL, while we improved the gas-tuning measurement by filtering a much narrower region  with the only help of the electrically tunable bandpass fiber filter described in SI Section~\ref{sec:setup}. 
We used a CW excitation power of 6~$\upmu$W (10~$\upmu$W) in the thermal (gas) tuning case. 
In both cases the data was acquired as raw time tags and is not background-corrected. 
The coincidence counts were evaluated with the software ETA~\cite{lin2021efficient} using a time binning of $200$~ps ($512$~ps) in the thermal (gas) tuning case. 

The G-center is known to feature three energy levels: a ground and excited singlet state and a metastable triplet state~\cite{udvarhelyi_identification_2021}. 
The metastable state introduces a bunching effect in the second-order correlation, and has been previously observed for G-centers~\cite{redjem_single_2020, hollenbach_engineering_2020}. This bunching effect is clearly visible in Fig.~\ref{sfig:g2s}b, while it is less evident in Fig.~\ref{sfig:g2s}a. 
We attribute this difference to the fact that different emitters have different electron trapping rates and other mesoscopic properties that dictate the bunching. 

\subsection{Tuning of cavity resonances}
\label{sec:Tuningofcavity}
Here we expand on the methods used to decouple the G-centers from our cavities. As mentioned in the main text, we performed both thermal and gas tuning to achieve decoupling. 

\subsubsection{Thermal tuning}
\label{sec:Thermaltuning}
\begin{figure}
  {\centering
  \includegraphics[width=0.53\textwidth]{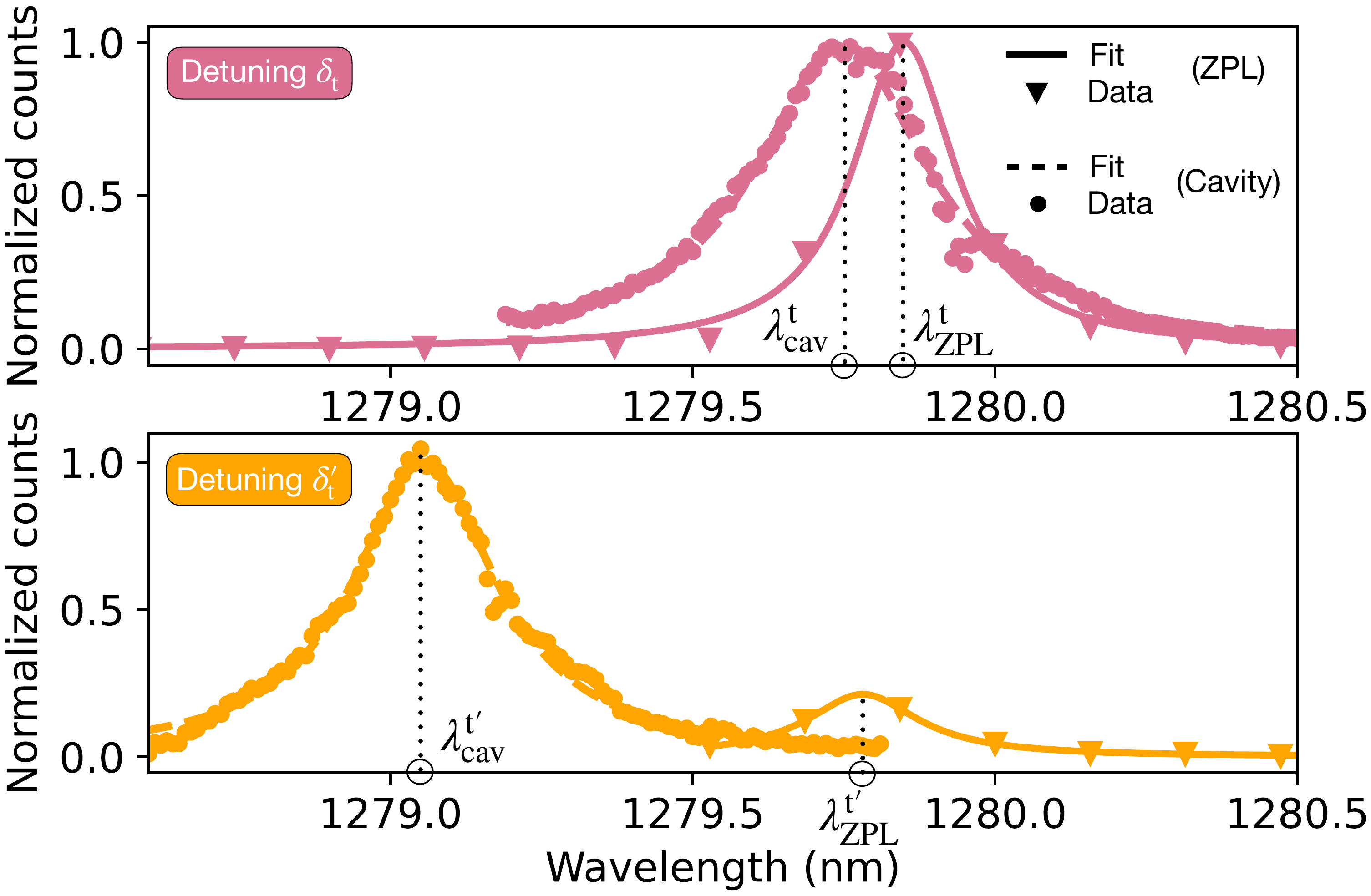}}
\caption{\textbf{Thermal tuning of a cavity.} Decoupling of the cavity-atom system is achieved by changing the temperature in the cryostat chamber from $\sim4$~K to $\sim24$~K, leading to a detuning of $\delta_{\mathrm{t}}$ (top) and  $\delta_{\mathrm{t}}^{\prime}$ (bottom), respectively. A significant cavity wavelength shift --- from $\lambda_{\mathrm{cav}}^{\mathrm{t}}$ to ${\lambda_{\mathrm{cav}}^{\mathrm{t}^{\prime}}}$ --- as well as a much less evident emitter ZPL shift --- from $\lambda_{\mathrm{ZPL}}^{\mathrm{t}}$ to ${\lambda_{\mathrm{ZPL}}^{\mathrm{t}^{\prime}}}$ --- are observed.}
\label{sfig:thermaltuning}
\end{figure}

Starting with a cryostat chamber temperature of $\sim4$~K, we define the detuning $\delta_{\mathrm{t}}$ as $\delta_{\mathrm{t}}=\lambda_{\mathrm{cav}}^{\mathrm{t}}-\lambda_{\mathrm{ZPL}}^{\mathrm{t}}$, where $\lambda_{\mathrm{ZPL}}^{\mathrm{t}}$ is the G-center ZPL wavelength and $\lambda_{\mathrm{cav}}^{\mathrm{t}}$ the cavity resonance wavelength. Fitting both the cavity and ZPL profiles to a Lorentzian function, we find the wavelengths to be $\lambda_{\mathrm{cav}}^{\mathrm{t}}= (1279.747  \pm 0.002)$~nm and $\lambda_{\mathrm{ZPL}}^{\mathrm{t}}= (1279.850 \pm  0.004)$~nm, and thus calculate $\delta_{\mathrm{t}}=(-0.103 \pm 0.004)$~nm. The cavity and ZPL profiles at $\sim4$~K are displayed in pink in Fig.~\ref{sfig:thermaltuning}. The extracted $Q$ factor of our cavity is $3725 \pm 50$.

In order to decouple the cavity-atom system, we warmed up the chamber to $\sim24$~K and observed the subsequent effect. As visible in orange in Fig.~\ref{sfig:thermaltuning}, the cavity shifted away spectrally from the ZPL, resulting in a reduction of the ZPL intensity. The cavity resonance is now found at $\lambda_{\mathrm{cav}}^{\mathrm{t}^{\prime}}= (1279.057  \pm 0.001)$~nm, and a less significant spectral shift is also observed in the ZPL wavelength, now at $\lambda_{\mathrm{ZPL}}^{\mathrm{t}^{\prime}}= (1279.781  \pm 0.001)$~nm. The latter effect is in line with what we recently reported in Ref.~\cite{prabhu_individually_2022}, where a non-volatile spectral shift in the ZPL of our single G-centers was observed. After the temperature variation, the detuning is found to be ${\delta^{\prime}_{\mathrm{t}}}=(-0.724 \pm 0.001)$~nm. Fig.~\ref{fig:coupling}d in the main text condenses this information by combining both detuning cases ${\delta_{\mathrm{t}}}$ and ${\delta^{\prime}_{\mathrm{t}}}$ in one plot. To highlight these detunings between the cavity and ZPL, we shifted the x-axis of the bottom plot in  Fig.~\ref{sfig:thermaltuning} by ($\lambda_{\mathrm{ZPL}}^{\mathrm{t}}$ - $\lambda_{\mathrm{ZPL}}^{\mathrm{t}^{\prime}}$~). In this way, the detunings between the cavity and ZPL are more clearly visible. However, because of this shift, it should be noted that the plot in Fig.~\ref{fig:coupling}d does not reflect the real difference between the cavity wavelengths before and after tuning. This information is preserved in Fig.~\ref{sfig:thermaltuning}.

All spectra were acquired with a grating density of $300$~gr/mm. The cavity reflectivity measurements were performed with our tunable CW narrowband IR laser while sweeping its wavelength and recording the cavity spectrum at each step. All spectra were then merged to obtain the complete cavity profile. 

\subsubsection{Gas tuning}
\label{sec:Gastuning}
To validate what we observed in the thermal tuning case, we spectrally shifted the cavity of a second cavity-atom system via a different tuning mechanism based on gas deposition and subsequent sublimation. 
In this case, the chamber temperature remained constant (at $\sim 8$~K) throughout our measurements. 
First, we coated our sample with a thin layer of gas (CO$_2$) by injecting the gas into our cryostat chamber via a dedicated gas line. 
The gas layer alters the mode index of the cavity and thus redshifts its wavelength. 
We assume first-order perturbation theory and follow the derivation in the SI of Ref.~\cite{panuski2022full}.
Using a refractive index for solid CO$_2$ of 1.4~\cite{warren_optical_1986}, and a representative mode for the photonic crystal cavity, we estimate a resonance shift of the order of $10$~nm for an infinite thickness of solid CO$_2$.

\begin{figure}
  {\centering
  \includegraphics[width=\linewidth]{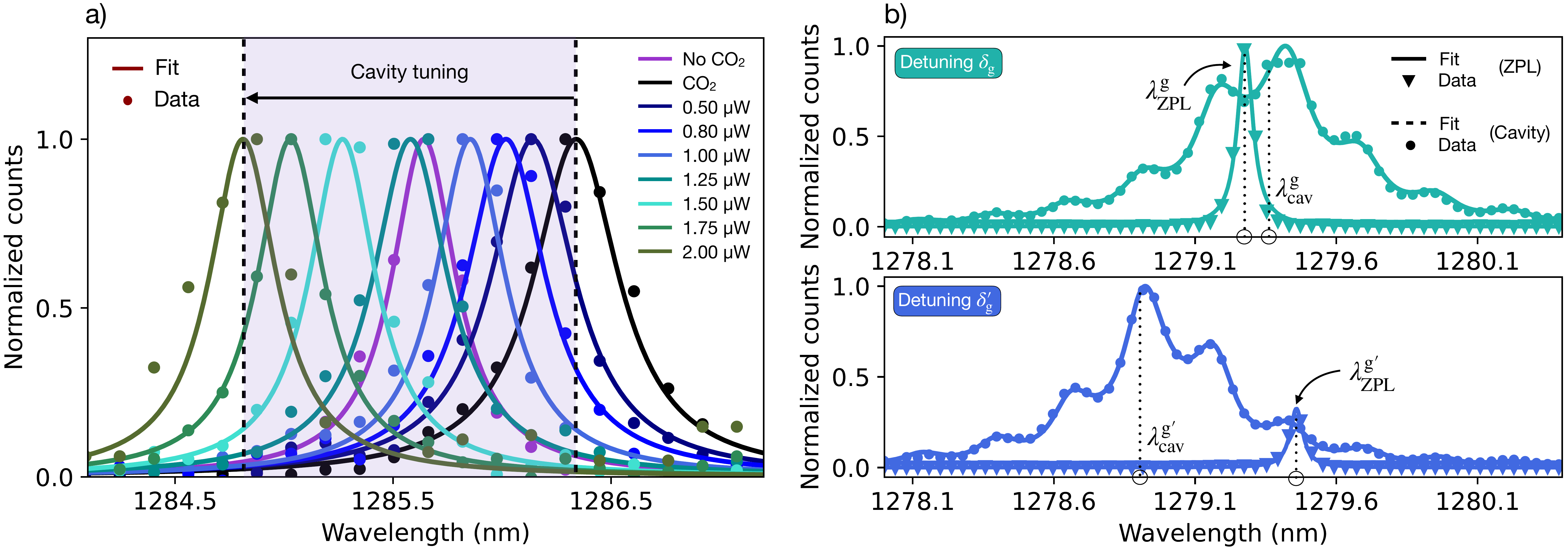}}
\caption{\textbf{In-situ gas tuning of a cavity.} a) Starting from its nominal wavelength (center of purple curve), the cavity resonance is redshifted by introducing CO$_2$, and controllably blueshifted by sequentially burning gas using a $532$~nm CW pump at different powers (indicated in the legend). b) Decoupling of the emitter from the cavity using this gas tuning method. The figure shows the ZPL and cavity profiles under different detunings $\delta_{\mathrm{g}}$ (before gas tuning, top) and $\delta_{\mathrm{g}}^{\prime}$ (after gas tuning, bottom).  A cavity wavelength shift --- from $\lambda_{\mathrm{cav}}^{\mathrm{g}}$ to ${\lambda_{\mathrm{cav}}^{\mathrm{g}^{\prime}}}$ --- as well as a ZPL shift --- from $\lambda_{\mathrm{ZPL}}^{\mathrm{g}}$ to ${\lambda_{\mathrm{ZPL}}^{\mathrm{g}^{\prime}}}$ --- are observed.}
\label{sfig:gastuning}
\end{figure}

We then performed gradual gas sublimation to achieve a controllable cavity resonance blueshift. Fig.~\ref{sfig:gastuning}a shows a representative experimental gas shifting of cavity spectra. Our maximum tuning, in the order of $1$~nm, aligns reasonably well with our theoretical estimate. Starting from the nominal cavity resonance wavelength (purple curve), we injected CO$_2$ and thus redshifted the cavity mode (black curve). We then illuminated the cavity with a $532$~nm CW pump for $\sim 0.5$~s at increasingly higher powers (displayed in the figure's legend), and thus achieved a controllable cavity resonance shift. We note that our cavity blueshifts further than its initial central wavelength, indicating that there is likely a significant leakage of CO$_2$ or other gases during the cooling, which we can sublimate using our optical pumping technique.

After depositing CO$_2$ on our sample, we acquired the ZPL and cavity spectra. We find their center wavelengths to be $\lambda_{\mathrm{ZPL}}^{\mathrm{g}} = (1279.277 \pm 0.001)$~nm and $\lambda_{\mathrm{cav}}^{\mathrm{g}}= (1279.354 \pm 0.002)$~nm, leading to a very small detuning of $\delta_{\mathrm{g}} = \lambda_{\mathrm{cav}}^{\mathrm{g}}-\lambda_{\mathrm{ZPL}}^{\mathrm{g}} = (0.077 \pm 0.002)$~nm as shown in green in Fig.~\ref{sfig:gastuning}b. Unlike the previous case, the cavity reflectivity measurements shown in Fig.~\ref{sfig:gastuning}b were performed with our superluminescent diode by simply illuminating the cavity and recording its reflectivity spectrum. Moreover, these spectra were acquired with a higher resolution compared to the previous experiments (here we used a grating density of $900$~gr/mm) which resulted in a slight wavelength offset likely due to calibration errors. This offset was taken into account when plotting the data, in order to enable a fair wavelength comparison between the thermal and gas tuning cases. 

While the value of $\lambda_{\mathrm{ZPL}}^{\mathrm{g}}$ was still obtained by fitting the ZPL profile to a Lorentzian function, the cavity spectrum required a different analysis due to the presence of parasitic oscillations in its profile, as visible in Fig.~\ref{sfig:gastuning}b. 
This behaviour was not observed in the thermal tuning case because of the lower resolution arising from a smaller grating density. 
It is known that these parasitic cavities are Fabry-P\'erot (FP) cavities arising from reflections in the silicon substrate~\cite{panuski2022full}. 
A FP cavity is described by an airy distribution~\cite{ismail_fabry-perot_2016} as
\begin{equation}
f_\text{FP} = \frac{A(1-R_1)^2R_2}{(1-\sqrt{R_1R_2})^2+4\sqrt{R_1R_2}\sin^2{\phi}},
\label{eq:airy}
\end{equation} 
with $A$ an amplitude fitting constant, $\phi=\pi (v-v_0)/v_\text{FSR}$, $v-v_0$ the optical frequency detuning, $v_\text{FSR}=c/2L$ the cavity free spectral range with $c$ the speed of light and $L$ the cavity length, and $R_1$ and $R_2$ the mirror reflectivities.

The photonic crystal cavity resonance is described by a Lorentzian as follows
\begin{equation}
f_\text{L}=\frac{B}{(\lambda - \lambda_0)^2 + (\Gamma/2)^2}+C,
\label{eq:lorentzian}
\end{equation} 
with $B$ and $C$ being fitting constants for the amplitude and offset, $\lambda$ and $\lambda_0$ the wavelength and wavelength offset, and $\Gamma$ the full width at half maximum (FWHM). We estimate the effect of the coupled system by using a product function
\begin{equation}
f_\text{tot}=f_\text{FP}f_\text{L}.
\label{eq:total}
\end{equation} 
We extract a cavity Q factor of $2136 \pm 30$. 

\begin{figure}
  {\centering
  \includegraphics[width=.48\linewidth]{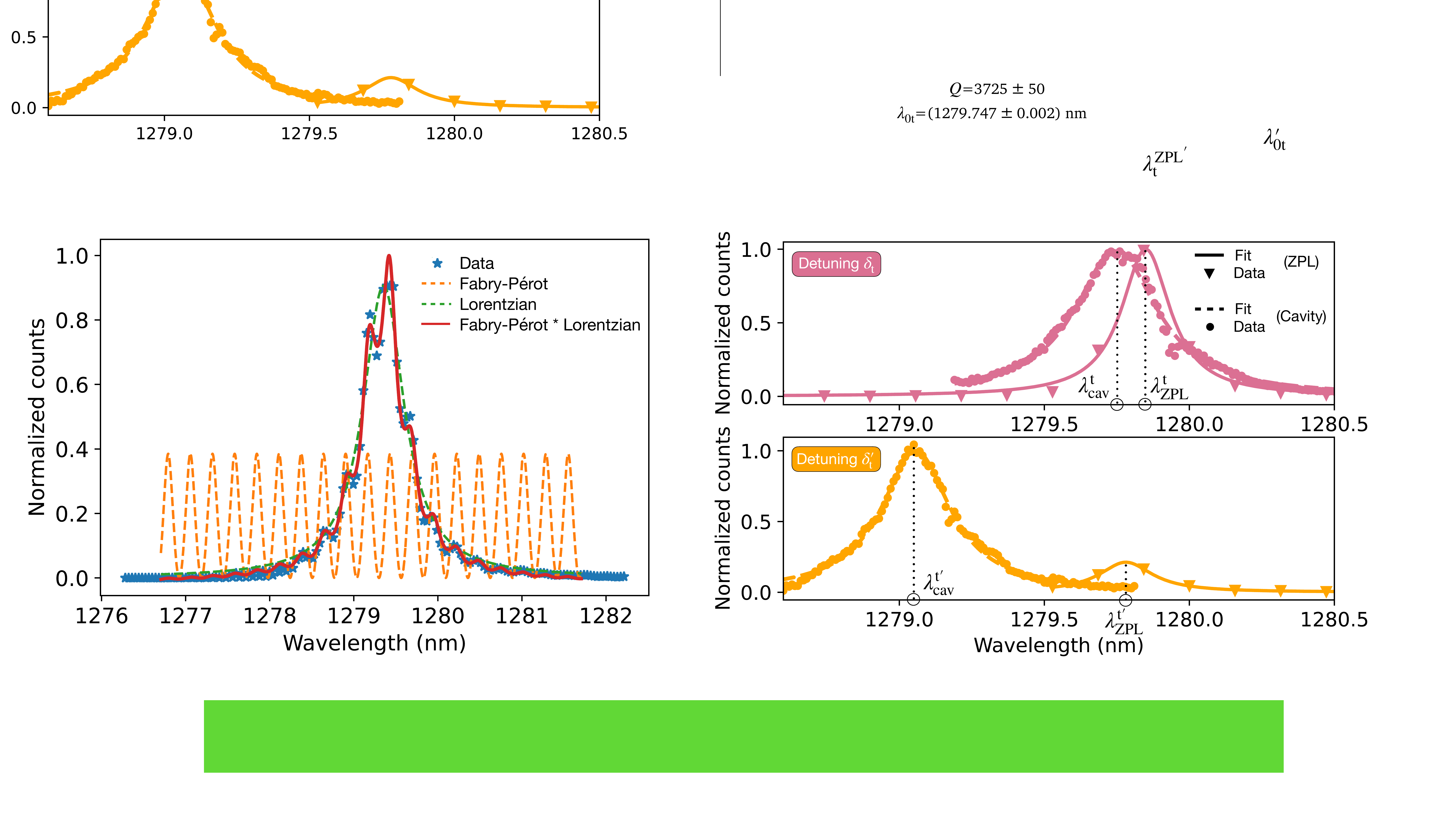}}
\caption{\textbf{Cavity resonance fit.} The combined Airy-Lorentzian (red) fits well our cavity reflectivity measurements (blue stars), and can be decomposed in a Fabry-P\'erot cavity (orange dashed) and a Lorentzian (green dashed).}
\label{sfig:cavityfit}
\end{figure}

To decouple our cavity-atom system, we illuminated a region in the vicinity of our cavity with $532$~nm CW laser light with powers up to 500~$\upmu$W. This resulted in a shift of our cavity resonance wavelength from  $\lambda_{\mathrm{cav}}^{\mathrm{g}}$ to $\lambda_{\mathrm{cav}}^{\mathrm{g}^{\prime}}= ( 1278.976 \pm 0.001)$~nm. We note that also in this case the optical pumping required for gas tuning introduces a non-volatile spectral shift in the ZPL of our single emitters, as reported in Ref.~\cite{prabhu_individually_2022}. The ZPL wavelength reads now $\lambda_{\mathrm{ZPL}}^{\mathrm{g}^{\prime}}=(1279.4587 \pm 0.0003)$~nm. A detuning of ${\delta^{\prime}_{\mathrm{g}}} = (- 0.483 \pm 0.001) $~nm is therefore achieved. These results are shown in blue in Fig.~\ref{sfig:gastuning}b. Considerations analogous to how the plot in Fig.~\ref{fig:coupling}d is realized (see previous subsection) hold for Fig.~\ref{fig:coupling}e as well.

\subsection{Derivation of quantum efficiency}
\label{sec:qe}

In the following, we derive the relevant rates from first principles and use them to estimate the G-center quantum efficiency 
of our system. We assume that 1) the cavity detuning does not significantly affect the Purcell factor of non-ZPL radiative phenomena (part of $\gamma_0$), and that 2) there is a negligible contribution of other radiative modes in our measurements due to our $0.1$~nm narrow filtering of the ZPL. Under these assumptions, we write the total emission rate $1/\tau$ and the collected photon flux $\phi$ for both the on- and off-resonance cases as

\begin{align}
\label{seq:rates}
& \quad \frac{1}{\tau_{\text{on}}}=\gamma_\text{R}+F_\text{P} \gamma_\text{R}+\gamma_\text{0},
  \\
\label{seq:rates1}
& \quad  \frac{1}{\tau_{\text{off}}}= \gamma_\text{R}+\gamma_\text{0},
   \\
\label{seq:rates2}
& \quad \phi_{\text{on}}=\eta (\gamma_{\text{R}}+ F_\text{P}\gamma_\text{R}),
  \\
\label{seq:rates3}
& \quad \phi_{\text{off}}=\eta \gamma_{\text{R}},
\end{align}

with $F_\text{P}$ the Purcell factor and $\eta=0.7$ the coupling efficiency into the wanted mode extracted from our simulations in Fig.~\ref{fig:cav_ff}. The quantum efficiency QE is defined as
\begin{equation}
\label{seq:qe}
\text{QE}=\frac{\gamma_\text{R}}{\gamma_\text{R}+\gamma_\text{0}}\frac{1}{F_\text{DW}},
\end{equation}
where $F_\text{DW}$ is the Debye-Waller factor, i.e. the fraction of the PL intensity emitted in the ZPL, and is extracted from the literature. Extracting $\gamma_{0}$ from the equation above and substituting it in the ratio between Eqs.~\ref{seq:rates} and ~\ref{seq:rates1}, that is $\tau_{\text{off}}/\tau_{\text{on}}$, we find an expression for $\tau_{\text{off}}/\tau_{\text{on}}$ as a function of QE:

\begin{equation}
\label{seq:tauqe}
\frac{\tau_{\text{off}}}{\tau_{\text{on}}}=1+F_{\text{P}} F_{\text{DW}}  \text{QE}.
\end{equation}

Following the procedure in Ref.~\cite{lefaucher2023cavity}, given that in our measurements the lifetimes remain constant within the error, we can define a bound on our QE by setting a threshold for the detection of the longest lifetime allowed by our off-resonance standard deviation $\tau_\text{off-th}=\tau_\text{off}-3\sigma_\text{off}$. As the on- and off-resonance lifetimes do not differ significantly, it holds that $\tau_{\text{on}}>\tau_{\text{off-th}}$. Therefore, we can derive an upper bound for the QE:

 \begin{equation} \label{eq:QEfinal}
 \text{QE}<\frac{\frac{\tau_\text{off}}{\tau_\text{off-th}}-1}{F_\text{P} F_\text{DW}}.
 \end{equation}

To calculate QE from Eq.~\ref{eq:QEfinal}, an estimate of the Purcell factor $F_{\text{P}}$ is needed. We can derive $F_{\text{P}}$ from the ratio between the photon fluxes $\phi_{\text{on}}$ and $\phi_{\text{off}}$ in Eqs.~\ref{seq:rates2} and~\ref{seq:rates3}:  
 \begin{equation}
 \label{eq:phionoff}
\frac{\phi_{\text{on}}}{ \phi_{\text{off}}}  = (1+F_{\text{P}}),
 \end{equation}
and thus find
 \begin{equation}
 \label{eq:purcell}
F_{\text{P}} = \frac{\phi_{\text{on}}}{ \phi_{\text{off}}} -1.
 \end{equation}

Deriving the Purcell factor $F_{\text{P}}$ from Eq.~\ref{eq:purcell} using our measured photon fluxes and substituting it in~\ref{eq:QEfinal}, we can estimate the QE to be bounded by $\sim 18 \%$ for a measured off-resonance lifetime value of $\tau_{\text{off}}=(6.09 \pm 0.25)$~ns.

\subsection{Comparison to the G-center literature}
\label{sec:comparison}

Table~\ref{tab:comparison} summarizes the reported literature on single emitters in silicon identified as G-centers in terms of their ZPL, estimated QE, $g^{(2)}(0)$, and excited state lifetimes.
We include a recent report on cavity-coupled ensembles~\cite{lefaucher2023cavity} and the simultaneous report on cavity-enhancement of singles~\cite{redjem_all-silicon_2023}, and use these to add cavity-enhanced excited state lifetime and $Q/V$ to the comparison. What reported in Table~\ref{tab:comparison} aligns with the hypothesis, originally brought forward by Ref.~\cite{baron_single_2022}, that the community is likely studying two different defects.
The first one, which we label $G_a$~\cite{hollenbach_engineering_2020, baron_single_2022, hollenbach_wafer-scale_2022, prabhu_individually_2022, komza_indistinguishable_2022, lefaucher2023cavity}, aligns with the original G-center ensembles, and features a ZPL at around $1278$~nm with a narrow inhomogeneous linewidth, a lower QE, and a short excited state lifetime.
The second one, $G_b$~\cite{redjem_single_2020, redjem_all-silicon_2023}, features a blue-shifted ZPL at $1270$~nm with a high QE and a longer excited-state lifetime.

Although our results suggest the possibility of two different artificial atom systems, the reported differences may still be due to a different host material, measurement setups, or fabrication protocols.

\setlength{\tabcolsep}{3.5pt}
\begin{table*}[!htbp]\centering
   \begin{tabular}{l r r r r r r l} 
	Reference	& ZPL  &	 QE	& $g^{(2)}(0)$ &	$\tau$ & $\tau^{\mathrm{cav}}$	& 	$Q/V_{\mathrm{cav}}$  & Notes\\ [0.5ex] 
	& (nm) &	($\%$)	&  & (ns) & (ns)	&  $(\lambda/n)^{-3}$ & \\ [0.5ex] 
	 \hline 
	 Redjem et al.~\cite{redjem_single_2020} & $1270.0\pm9.1$ & $\sim50$* & 0.3 &$35.8\pm0.2$ & & & 29 emitters \\ 
	 Hollenbach et al.~\cite{hollenbach_engineering_2020} & $1278.3\pm0.1$ & & $0.07\pm0.04^\dagger$ & $3.8$ & & & 12~\textmu m SOI, 12 emitters\\ 
	 Baron et al.~\cite{baron_single_2022} & $1279$ & & $\sim0.3$ & $4.5$ & & & 1 emitter measured \\ 
	 Hollenbach et al.~\cite{hollenbach_wafer-scale_2022} & $1278$ & & $0.36\pm0.06$ & $10.0$ & & & 1 emitter measured \\ 
	 Prabhu et al.~\cite{prabhu_individually_2022} & $1278.7\pm1.1$ & >1* & $0.38\pm0.08$ & $8.3\pm0.7$ & & & Waveguide, 37 emitters \\ 
	 Komza et al.~\cite{komza_indistinguishable_2022} & $1278$ & >2* & $0.15\pm0.02$ & $4.6$ & & & Waveguide, 1 emitter \\ 
	 Redjem et al.~\cite{redjem_all-silicon_2023} & $1275$ & & $0.30\pm0.07$ & $33.3$ & $6.7$ & 4862 & L3 cavity, 1 emitter \\ 
	 Lefaucher et al.~\cite{lefaucher2023cavity} & $1279$ & <10** &  & $5.6 \pm 0.1$ & $5.6 \pm 0.1$ & 417 & Ensembles, static rings \\ 
	 \textbf{This work} & $1279$ & <18& $0.03\pm \substack{0.07 \\ 0.03}$ & $6.1 \pm 0.3$ & $5.50 \pm 0.05$ &  $>3725$ & Opt. L3 cavity, 1 emitter \\ 
\end{tabular}
  \caption{Comparison of measured properties for the reported G-centers in the literature. All the listed references other than Lefaucher et al.~\cite{lefaucher2023cavity} report single-photon emitters. If several emitters are listed in Notes, the reported error is the statistical standard deviation. Otherwise it is the measurement/fit error. *Estimated from photon counts and setup loss. **Extracted from ensembles and static cavities. $\dagger$ Background corrected.}
\label{tab:comparison}
\end{table*}

\appendix*

\end{document}